\documentclass[twocolumn,3p,number,sort&compress]{elsarticle}
\usepackage{lineno} 
\usepackage{mathtools} \usepackage{graphicx}
\usepackage{subfig}

\journal{Nuc. Instr. and Meth. A}

\begin{document} 

\begin{frontmatter}

\title{Upgraded photon calorimeter with integrating readout for Hall A Compton
Polarimeter at Jefferson Lab}

\author[ad1]{M. Friend\corref{cor}} \ead{mfriend@andrew.cmu.edu}
\author[ad1,ad2]{D.  Parno} \author[ad1,ad3]{F. Benmokhtar} \author[ad4]{A.
Camsonne} \author[ad5]{M. Dalton} \author[ad1]{G. B.  Franklin} \author[ad1]{V.
Mamyan} \author[ad4]{R. Michaels} \author[ad4]{S. Nanda} \author[ad5]{V.
Nelyubin} \author[ad5]{K. Paschke} \author[ad1]{B.  Quinn} \author[ad6]{A.
Rakhman} \author[ad6]{P. Souder} \author[ad5]{A. Tobias} 

\address[ad1]{Carnegie Mellon University, Department of Physics, 5000 Forbes
Ave, Pittsburgh, PA 15213, USA} \address[ad2]{University of Washington, Center
for Experimental Nuclear Physics and Astrophysics and Department of Physics,
Seattle, WA 98195, USA} \address[ad3]{Christopher Newport University, Department
of Physics, Computer Science and Engineering, 1 University Place, Newport News,
VA 23606, USA} \address[ad4]{Thomas Jefferson National Accelerator Facility,
12000 Jefferson Ave, Newport News, VA 23606, USA} \address[ad5]{University of
Virginia, Department of Physics, 382 McCormick Rd, Charlottesville, VA 22904,
USA} \address[ad6]{Syracuse University, Department of Physics, Syracuse, NY
13244, USA}

\cortext[cor]{Corresponding author. Tel: +1-412-268-6949, fax: +1-412-681-0648}

\begin{abstract}

The photon arm of the Compton polarimeter in Hall A of Jefferson Lab has been
upgraded to allow for electron beam polarization measurements with better than
1\% accuracy.  The data acquisition system (DAQ) now includes an integrating
mode, which eliminates several systematic uncertainties inherent in the original
counting-DAQ setup.  The photon calorimeter has been replaced with a Ce-doped
Gd\(_2\)SiO\(_5\) crystal, which has a bright output and fast response, and
works well for measurements using the new integrating method at electron beam
energies from 1 to 6~GeV.

\end{abstract}


\end{frontmatter}

\section{Introduction} 

The Jefferson Lab Continuous Electron Beam Accelerator Facility (CEBAF)
\cite{Leemann:CEBAF01} delivers up to 200~\(\mu\)A of 1-6~GeV highly polarized
electrons to three experimental halls.  A Compton backscattering polarimeter was
installed in Hall A in 1999, to continuously monitor the electron beam
polarization during experimental data-taking \cite{Escoffier:ComptonEBP05}.  

The longitudinal polarization of an electron beam can be defined as
\begin{equation}\label{eq:Pel} P_e = \frac{N_e^+ - N_e^-}{N_e^+ + N_e^-},
\end{equation} where \(N_e^{+(-)}\) is the number of electrons with positive
(negative) helicity in a single accelerator helicity state.  At Jefferson Lab,
the electron helicity state is flipped rapidly during normal operations, as
described in Sec.\ \ref{sec:Apparatus}.

In the Hall A Compton polarimeter, the longitudinally polarized electron beam is
allowed to scatter from circularly polarized laser light in a high-finesse
Fabry-P\'{e}rot cavity \cite{Jorda:ComptonCavity98}. The polarimeter takes
advantage of the fact that the Compton scattering cross-sections depend on the
relative polarizations of the incident electrons and photons, and these
cross-sections are very well known \cite{Lipps:ComptonPol54i}. One can measure a
Compton scattering asymmetry \begin{equation}\label{eq:AsymScatt} A_{exp} =
\frac{S^{+} - S^{-}}{S^{+} + S^{-}}, \end{equation} where \(S^{+(-)}\) is the
scattered photon signal for positive (negative) helicity electrons, and can be
integrated or counted.  At the Compton endpoint, \(A_{exp}\) is positive for
left-circularly-polarized photons.  Given ideally (100\%) polarized electron and
photon sources, \(A_{exp}\) would be equal to a calculable \(A_{th}\), the
theoretical Compton scattering asymmetry derived by combining the calculated
spin-dependent Compton cross-section with the appropriate experimental response
function \cite{Prescott:ComptonSDC73}.  

If the polarization is the same for both electron beam helicity states,
\begin{equation}\label{eq:AMeas} A_{exp} = P_eP_{\gamma}A_{th},\end{equation}
where \({P}_e\) is the electron beam polarization and \(P_{\gamma}\) is the
photon beam polarization, and thus \(P_e = A_{exp}/P_{\gamma}A_{th}\).  The
experimental asymmetry is measured over pairs of helicity windows, and if
\(P_e\) is not identical for the two helicity states, the asymmetry is given by
\begin{equation}A_{exp} =
\bar{P}_{e}P_{\gamma}A_{th}\frac{1}{1+\frac{1}{2}(P^+_e-P^-_e)A_{th}},
\end{equation} where \(P^{+(-)}_e\) is the electron polarization for a positive
(negative) helicity accelerator state and \(\bar{P}_{e}\) is the average of the
two.  If the asymmetry \(A_{th}\) is sufficiently small, as in past Hall A
experiments such as HAPPEX-III \cite{HAPPEXiiiProposal}, the correction due to
the final factor in Eq.\ \ref{eq:AMeas} is negligible and \(A_{exp} \simeq
P_{e}P_{\gamma}A_{th}\); however, for large enough \(A_{th}\), \(P^+_e-P^-_e\)
could play a significant role.  The measured scattering asymmetry for
longitudinally polarized electrons from circularly polarized photons thus gives
a sensitive measurement of the electron beam polarization, given accurate
knowledge of \(A_{th}\) and \(P_{\gamma}\).

The theoretical asymmetry, \(A_{th}\), is calculated using a GEANT4 simulation
\cite{Agostinelli:Geant4_03}, which is customized for the specific kinematics,
apparatus, and integrating or counting measurement method chosen
\cite{Parno:th11}, and includes a corresponding radiative correction
\cite{Denner:ComptonRadiative99}.  A detailed discussion of this simulation will
be given in a future publication.

This paper will discuss the Compton polarimetry apparatus in Hall A, including
details about the new photon detector, and the upgraded DAQ.  Analysis details
will also be given, including a discussion of systematic errors.  Finally,
electron beam polarization results from the HAPPEX-III experiment will be given.

\section{Apparatus} \label{sec:Apparatus}

The Jefferson Lab electron beam source is a strained superlattice GaAs
photocathode illuminated by a circularly polarized laser.  The electron beam
helicity is controlled by a Pockels Cell (PC) which sets the polarization
direction of the source laser; the polarity of the voltage setting on the PC is
reversed to flip the helicity.  This allows for a fast helicity flip rate: the
electron beam helicity is typically gated in windows of anywhere from 1/30~s to
1/1000~s.  The helicity is typically unstable for a time period on the order of
\({\sim}\)100~\(\mu\)s (\(T_{settle}\)) after the PC voltage is changed, and is
then stable for the remainder of the helicity period (\(T_{stable}\)) (as
discussed in Sec.\ \ref{sec:FADC}).  For each experiment, the helicity for
sequential windows is set in groups, e.g.\ quartets (\(+{-}{-}+\) or
\(-{+}{+}-\)), where a pseudo-random number generator is used to determine the
order of the groups.  To eliminate potential systematic effects, the signal
which tells the Hall A electronics the electron beam helicity may be delayed at
the source by some set number of helicity windows (usually 8).

\begin{figure}[] \caption{Design of the Hall A Compton polarimeter.  The angles
are exaggerated: the crossing angle between the electron and photon beams is
23~mrad.  \label{fig:comptpol}} \begin{center}
\includegraphics[width=8.0cm]{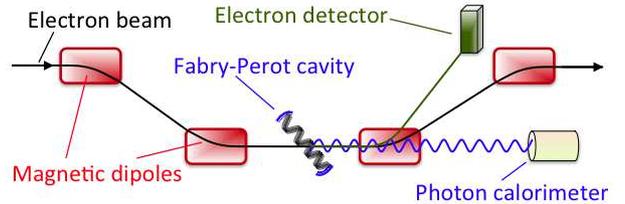} \end{center}
\end{figure}

A schematic of the Hall A Compton polarimeter is shown in Fig.\
\ref{fig:comptpol}.  In the entrance tunnel to Hall A, the electron beam is bent
downward by the first two magnetic dipoles of the Compton chicane to a parallel
path, where the electrons scatter off of laser photons in the Fabry-P\'{e}rot
cavity.  Backscattered photons are detected as gamma rays in the photon
detector.  Scattered electrons are separated from unscattered ones in the third
dipole, and can be detected in a silicon microstrip electron detector
\cite{Nanda:ComptonUpgrade04}.  Approximately one electron in \(10^9\) scatters;
the remainder are bent back to horizontal in the fourth dipole, and continue to
the Hall A fixed target, allowing for continuous polarization measurement
without significantly disturbing the incident electron beam.  

The Compton photon source, which was upgraded from infrared (\(\lambda =
1064\)~nm) to green (\(\lambda = 532\)~nm) in 2010, provides linearly polarized
laser light which is transformed to circular polarization using a
remotely-controlled quarter-wave plate (QWP).  The QWP is rotated by 90 degrees
to switch between left- and right-circularly polarized laser light.  The laser
light is locked in resonance in the Fabry-P\'{e}rot cavity for \({\sim}90\)~s,
then the cavity is taken out of resonance while the QWP is rotated to switch
polarizations.  The out-of-resonance period lasts \({\sim}30\)~s and is used for
background determination, since, when the cavity is unlocked, there is
negligible photon power at the Compton interaction point (CIP).  The main
background sources are beam-halo bremsstrahlung and synchrotron radiation.

The laser photon polarization is monitored on-line at the cavity exit by two
powermeters, which measure the outputs of a Wollaston prism polarizing beam
splitter.  A QWP lies before the Wollaston prism.  About once per day, the angle
of this QWP is scanned and the corresponding power meter measurements are used
to determine the laser polarization at the cavity exit. These measurements, in
combination with a polarization transfer function from off-line absolute
polarization measurements at the CIP and cavity exit, are used to determine the
laser polarization state at the CIP, \(P_{\gamma}\) \cite{Baylac:th00}.  The
photon polarization is \({\sim}99\%\).

The scattered photons are detected in a newly installed Ce-doped
\(\text{Gd}_2\text{SiO}_5\) (GSO) photon calorimeter, discussed in Sec.\
\ref{sec:GSO}.  The scintillation photons are detected by a single
photomultplier tube (PMT), and the resulting signals are integrated with a
customized Flash Analog-to-Digital Converter (FADC), discussed in Sec.\
\ref{sec:FADC}.  The upgraded FADC DAQ allows for a photon-arm-only polarization
measurement.

\section{Integrating vs.\ Counting Modes} \label{sec:intvscount}

The Compton photon scattering asymmetry may be measured either by counting the
number of scattered photons detected for each helicity state (determined by
counting the number of detected PMT pulses which cross a chosen discriminator
threshold), or by integrating the scattered photon signal for each helicity
state (determined by integrating all of the charge collected from the PMT for
each helicity state).  An asymmetry over helicity states, as in Eq.\
\ref{eq:AsymScatt}, may then be calculated using either value.

\begin{figure}[] \begin{center}
\includegraphics[width=8.0cm]{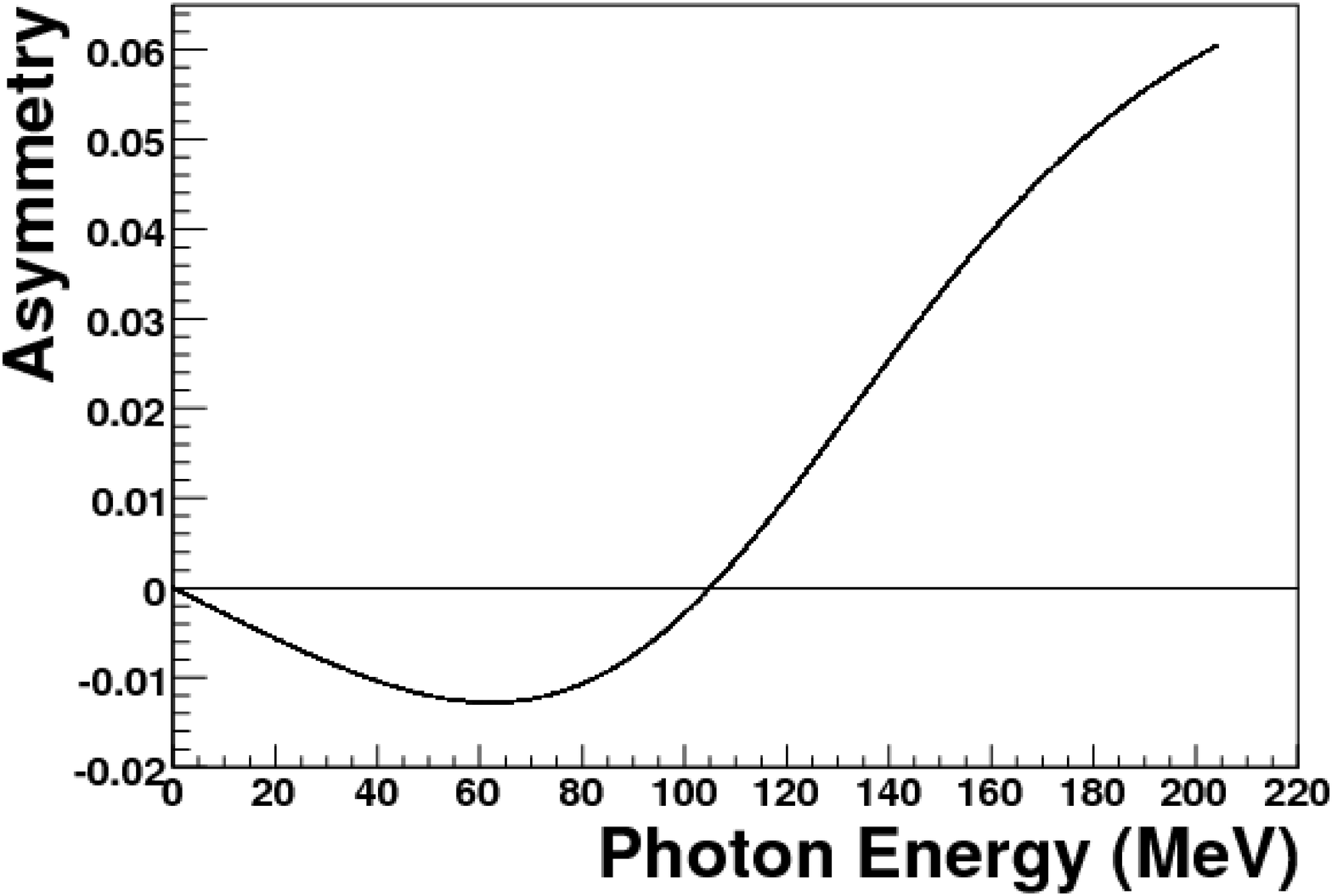} \caption{Theoretical
Compton scattering asymmetry as a function of scattered photon energy for an
electron beam energy of 3.4~GeV and a photon beam wavelength of 1064 nm.  The
maximum scattered photon energy for these kinematics is 204~MeV.
\label{fig:AsymVsEng}} \end{center} \end{figure}

Fig.\ \ref{fig:AsymVsEng} shows the theoretical Compton asymmetry, \(A_{th}\),
plotted as a function of scattered photon energy for a 3.4-GeV electron beam and
a photon wavelength of 1064 nm.  A measurement weighted by the energy deposited
by each Compton photon emphasizes the large-positive-asymmetry part of the curve
while suppressing the negative-asymmetry part, and therefore enhances the
measured asymmetry compared to simply counting the number of Compton photons
detected for each helicity.  Integration of the scattered photon signal yields
an energy-weighted measurement.

There are several systematic uncertainties inherent in making either a counting
or an integrating Compton polarization measurement, and these are considered
below.  The upgraded integrating DAQ design, however, eliminates several sources
of systematic error from the original counting DAQ \cite{Baylac:ComptonEBP02},
and by carefully controlling the integrating DAQ systematics, a very precise
measurement may be made.

Integration of the scattered-photon signal increases sensitivity to detector
non-linearities, since any systematic distortion of the detected
energy-deposited spectrum will also systematically distort the energy-weighted
asymmetry.  Integration also decreases the signal-to-background ratio compared
to counting events above a discriminator threshold, which would essentially
eliminate low-energy synchrotron radiation background.  Sensitivities to
pedestal fluctuations and scintillator afterglow are also increased.  Detector
non-linearities and scintillator afterglow can be controlled with careful
detector design and study.  Enhanced background and pedestal-fluctuation
sensitivities increase the statistical error bar of the measurement, but are not
a cause of systematic error, as long as background subtraction is done properly.
Since the measurement is not statistics-limited, this increased statistical
error is not a significant problem.

Signal integration, however, also eliminates sensitivities to threshold,
pileup, and dead-time effects inherent in a counting measurement, which would
be a potentially major source of systematic uncertainty.  A no-threshold
measurement also eliminates the need for precise calibration of the Compton
photon spectrum (e.g.\ with the scattered electron energies measured by the
Compton electron detector), which is required for the counting mode DAQ.  This
allows a precise stand-alone photon measurement to be made with the new
integrating Compton DAQ.

\section{Photon Detector} \label{sec:GSO}

\begin{figure}[] \begin{center} \includegraphics[width=8.0cm]{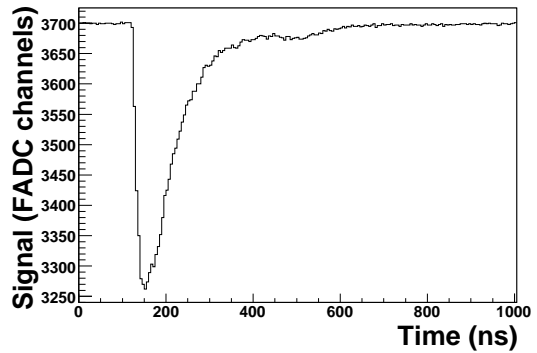}
\caption{A GSO signal from an incident photon of \({\sim}\)100~MeV.
``Snapshot'' taken using the triggered mode of the DAQ.  The signal has a
\({\sim}\)1.1 V peak which corresponds to \({\sim}\)10000 summed raw-ADC units
(raus).  Sampled voltage is reflected by a decrease from pedestal value.}
\label{fig:trigpulse} \end{center} \end{figure}

The new photon detector is a cylindrical GSO crystal that is 15~cm long and has
a 6-cm diameter.  It sits on a remotely-controllable table which can be moved
horizontally and vertically; this is useful both for detector alignment and for
moving the detector out of the beam when the background is very high.  It has
replaced the original photon detector, a 5\(\times\)5 array of 2~cm \(\times\)
2~cm \(\times\) 23~cm \(\text{PbWO}_4\) crystals
\cite{Neyret:ComptonCalorimeter00}.  GSO has a fast and bright signal: it
produces \({\sim}450\) optical photons per MeV (electron equivalent), with a
stable signal width of \({\sim}85\)~ns full width at half maximum.  A typical
photon pulse is shown in Fig.\ \ref{fig:trigpulse}.

The Hall A Compton photon beamline has also been modified, including the
installation of a new collimator directly upstream of the photon detector.  This
collimator, which reduces bremsstrahlung background and defines the Compton
angular acceptance, is a 5-cm-thick, 8-cm-diameter lead cylinder with a manually
interchangeable aperture of 2~cm maximum diameter, located \({\sim}\)6~m
downstream of the CIP and \({\sim}\)10~cm upstream of the GSO face.

A thin (0.25 to 8~mm thick, interchangeable), 4-cm-diameter lead disk is mounted
on the downstream side of the collimator.  This lead disk serves to shield the
photon detector from low-energy synchrotron radiation background passing through
the collimator aperture; the installed disk is chosen to be as thin as possible
to achieve acceptable background rates, since the lead also stops low-energy
Compton photons and introduces a distortion of the Compton spectrum which must
be accounted for when calculating the experimentally relevant \(A_{th}\).

\subsection{Afterglow}

Scintillator afterglow can affect the results of a Compton measurement, since
long-term afterglow might last for multiple helicity cycles, or could even
continue while the laser cavity is not locked and the background measurement is
being made.  The use of a scintillator that doesn't have a long-term afterglow
is therefore vital.  

Douraghy et al.\ have measured a \({\sim}\)5-\(\mu\)s afterglow in GSO
\cite{Douraghy:ScintAfterglow06}, which is short compared to a 33-ms helicity
window.  Compton polarimeter data were also studied to confirm that any
afterglow effect is negligible in this setup.  A detector response due to
afterglow would manifest as integrated signal that slowly decreases after beam
trips (instead of falling immediately to zero) or as an increased integrated
signal in those helicity windows which follow higher-rate helicity windows (an
increase in signal in a window following an electron-photon helicity parallel
window vs.\ an antiparallel one).  No afterglow effect is distinguishable from
statistical fluctuations in these studies, demonstrating that, in this instance,
these effects are negligible.

\subsection{Detector and Collimator Positioning}\label{sec:positioning}

Kinematics relate photon energy and scattering angle for backscattered Compton
photons: \begin{equation}k'=\frac{4kE^2}{4kE+m^2+\theta_{\gamma}^2E^2}
\end{equation} in the lab frame, where \(k\) and \(k'\) are the initial and
scattered photon energies respectively, \(E\) is the incident electron energy,
\(m\) is the electron mass, and \(\theta_{\gamma}\) is the scattered photon
angle.  Lower-energy scattered photons have a larger deviation from the initial
electron direction.  Ideally, the collimator intercepts only the lowest-energy
Compton photons; any mis-centering of the collimator causes some fraction of
higher-energy photons to be intercepted.  This mis-alignment must at the least
be properly reflected in the GEANT4 simulation used to calculate \(A_{th}\), and
should be minimized, if possible.  A mis-centered positioning of the GSO
calorimeter may also result in photon-energy-dependent changes in
energy-weighting which should also be included in \(A_{th}\), although this
effect is considerably less sensitive to small offsets than is the collimator
position.  Both effects can be minimized if the position of the center of the
backscattered photon beam can be accurately determined (and the collimator and
detector can be centered on it, if possible).

\begin{figure*}[t] \begin{center} \includegraphics[bb=0 0 1100 420,clip=true,
width=14.0cm]{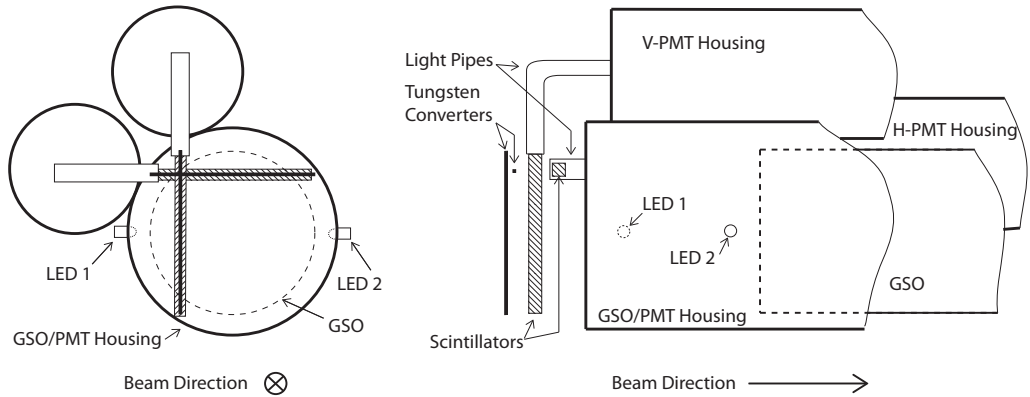} \caption{Front and side view of the GSO
detector housing including the tungsten converters and scintillators used for
determining the photon beam position.  Here the V-PMT and H-PMT read out the
vertical and horizontal converters respectively.  The center of each converter
bar is 20 mm away from the center of the GSO.  LEDs placed in front of the
detector, for use with the detector linearity test rig described in Sec.\
\ref{sec:linTesting}, are also shown. \label{fig:fingers}} \end{center}
\end{figure*}

A position monitor has therefore been implemented: two 0.1~cm \(\times\) 0.1~cm
\(\times\) 4.1~cm pieces of tungsten have been precisely positioned and bolted
to the GSO housing upstream of the front face of the crystal, as shown in Fig.\
\ref{fig:fingers}.  One tungsten converter is placed horizontally, positioned
20~mm above the center of the GSO crystal; the other is placed vertically and is
positioned 20~mm towards beam left from the center of the crystal.  During
normal data-taking, these lie out of the path of the Compton photons.  The
tungsten pieces sit in front of small bars of scintillator, which are read out
by PMTs.  When the tungsten converters intercept the photon beam they initiate
showers, which are then detected by the scintillators.

Using the remotely controlled photon detector table, the position of which can
be precisely set and read-back electronically to 0.2 mm, the detector may be
moved while the electron beam is in the hall.  The entire photon detector is
scanned horizontally while centered vertically, and then scanned vertically
while centered horizontally, and the trigger rate in the PMTs reading out each
scintillator bar is monitored.  When the tungsten converter crosses the
scattered photon beam, the trigger rate increases notably.  These tungsten
converters can therefore be used to precisely determine the location of the
scattered photon beam relative to the center of the GSO, and the photon detector
table can be positioned accordingly.

The photon detector also has an attached precision-placed long metal
``pointer'', which moves relative to a grid mounted on the side of the
stationary collimator.  This can be used to precisely determine the position of
the photon beam at the collimator; the collimator can then be positioned
manually so that the photon beam is centered on the collimator hole with a
precision of \({\sim}\)0.5 mm.  

\subsection{Detector Linearity Testing} \label{sec:linTesting}

Since an integrating asymmetry measurement is especially sensitive to detector
non-linearities, as described in Sec.\ \ref{sec:intvscount}, a test rig
\cite{Friend:Pulser11} using light-emitting diodes (LEDs) has been designed and
built in order to accurately determine the response of the PMT used to read out
the GSO photon response pulses.  This test rig also allowed the design of the
PMT base to be fine-tuned to achieve good linearity and rate stability.

The precise results of the linearity measurement made using this device are an
input into the GEANT4 simulation used to calculate \(A_{th}\).

The LED test rig can also be used to monitor PMT rate-dependent gain shifts.
For example, there is an observed 1\% increase in gain for data taken in the
presence of Compton signal plus background compared to that taken in the
presence of just background during the HAPPEX-III experiment.  This was measured
by flashing an LED at a range of stable brightnesses while locking and unlocking
the photon Fabry-P\'{e}rot cavity, and triggering the DAQ on the same trigger as
the LED.  After taking pileup effects into account, any systematic difference in
LED pulse size between cavity-locked and cavity-unlocked states is due to a
detector gain shift.  This effect can then be accounted for during analysis.

\section{Data Acquisition System} \label{sec:FADC}

The upgraded integrating data acquisition system is based upon the 200 MHz
Struck SIS3320 12-bit VME FADC.  The FADC continuously samples the photon
detector PMT output at 200 MHz (using an external 40 MHz clock internally
converted to 200 MHz).  The FADC sums the sampled data into six 36-bit
accumulators, which sum ADC values between an external \(T_{start}\) and
\(T_{stop}\) signal.  This accumulator mode is implemented through a
customization of the FADC, as described in Sec.\ \ref{sec:Accums}.
Simultaneously, the FADC stores all of the samples for a single helicity window
as sequential entries in one of two internal buffers.  The buffer is switched
after each helicity window.  A selected number of samples of the stored data can
be read out for each of a limited number of triggers in triggered-mode running,
as described in Sec.\ \ref{sec:TriggeredMode}.  

Necessary diagnostic signals, such as readback from beam current and position
monitors in the Compton beamline and the measured power of photons transmitted
through the Compton cavity, are converted to frequency signals in a
voltage-to-frequency converter, and are sent to two CAEN V560 scalers.  The
scalers are read out every helicity window, so that cuts on these quantities can
be made on a window-by-window basis during data analysis.  Other quantities,
such as trigger rates and clock pulses, are also counted by these scalers.  

\begin{figure}[] \begin{center} \includegraphics[bb=145 530 378 710,clip=true,
width=7.0cm]{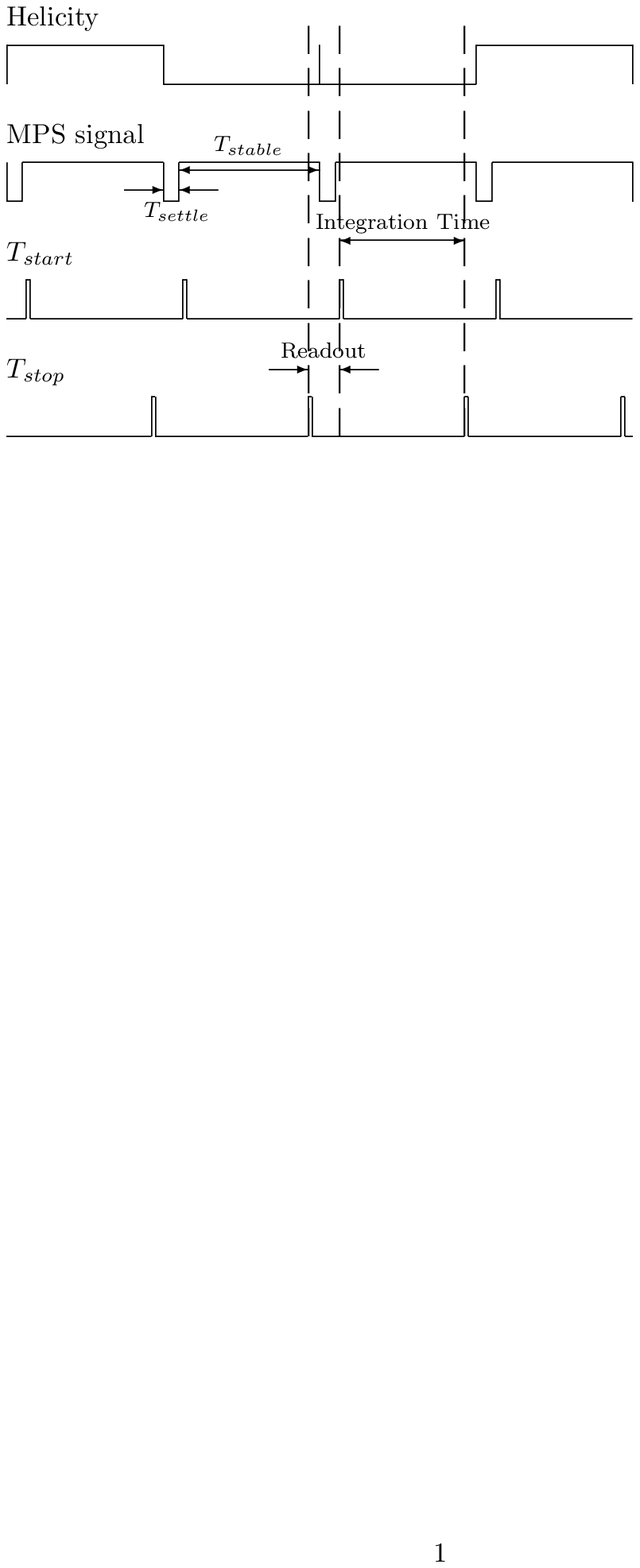} \caption{Timing structure for helicity windows.  The
timing signals are not drawn to scale, since the values of \(T_{settle}\),
\(T_{stable}\), and the integration time are selectable. \label{fig:MPSfig}}
\end{center} \end{figure}

The timing structure for the DAQ is shown in Fig.\ \ref{fig:MPSfig}.  The
external \(T_{start}\) and \(T_{stop}\) signals are generated using a customized
VME Timing Board \cite{Miller:th01}.  This VME module takes the accelerator
helicity timing signal (called the MPS signal) as a TTL input, outputs the
\(T_{start}\) signal a programmable interval of at least 15~\(\mu\)s later, and
outputs the \(T_{stop}\) signal a programmed interval after the \(T_{start}\)
signal.  The time between the \(T_{start}\) and \(T_{stop}\) signals must be set
to less than the length of the accelerator's helicity window.  Data readout is
initiated by the \(T_{stop}\) signal, and the time period between \(T_{stop}\)
and the next \(T_{start}\) is used to read out the accumulator and scaler data.
The triggered data can be read out during the following helicity window, since
the samples for two adjacent windows are stored in two separate buffers.  Since
the time between \(T_{start}\) and \(T_{stop}\) is completely programmable, the
integrating DAQ can be run at any accelerator helicity-flip period.  

\subsection{Accumulator Mode} \label{sec:Accums}

\begin{figure*}[t] \begin{center} \includegraphics[width=13.0cm]{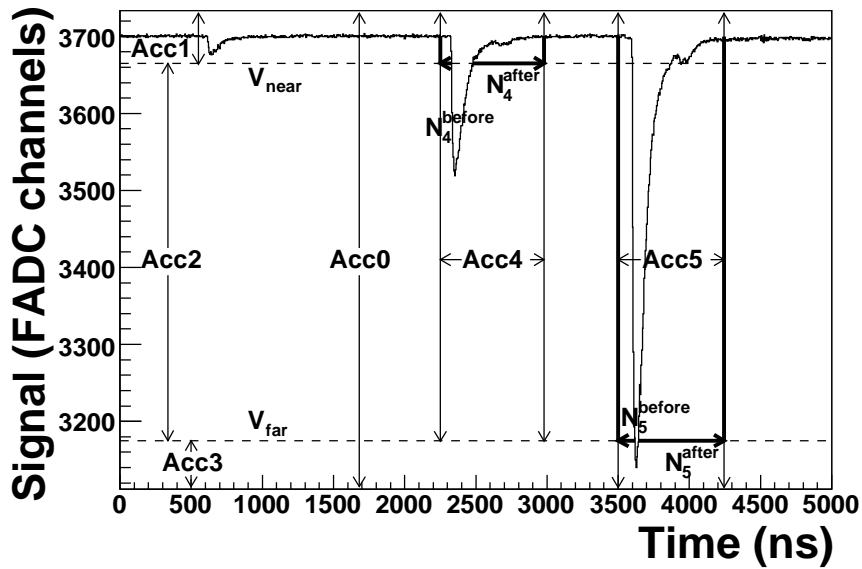}
\caption{The six accumulators are represented schematically.  Signal size within
a sample is represented by a decrease in value from a pedestal value.
\label{fig:accs}} \end{center} \end{figure*}

There are six accumulators, represented in Fig.\ \ref{fig:accs}, which are read
out for each helicity window:

\begin{itemize} \item Accumulator 0 (\emph{All}) sums all samples between the
external \(T_{start}\) and \(T_{stop}\) signals.  

\item Accumulator 1 (\emph{Near}) sums all samples that fall closer to the
pedestal than a threshold near the pedestal, \(V_{near}\) (low-energy photons).

\item Accumulator 2 (\emph{Window}) sums all samples between \(V_{near}\) and a
threshold far from the pedestal, \(V_{far}\) (with \(V_{near}\) between the
pedestal and \(V_{far}\)).

\item Accumulator 3 (\emph{Far}) sums all samples beyond \(V_{far}\) (the tips
of pulses from high-energy photons past the Compton edge).

\item Accumulator 4 (\emph{Stretched Window}) sums starting a set number of
samples, \(N_4^{\mathit{before}}\), before the signal crosses \(V_{near}\), and
continues to integrate until another set number of samples,
\(N_4^{\mathit{after}}\), after the signal crosses \(V_{near}\) again, except
that samples which are included in accumulator 5 are not included in accumulator
4.

\item Accumulator 5 (\emph{Stretched Far}) sums starting a set number of
samples, \(N_5^{\mathit{before}}\), before the signal crosses \(V_{far}\), and
continues to integrate until another set number of samples,
\(N_5^{\mathit{after}}\), after the signal crosses \(V_{far}\) again.
\end{itemize}

The \emph{All}, \emph{Window}, and \emph{Stretched Window} accumulators are
intended as possible measures of Compton signal, while the \emph{Near},
\emph{Far}, and \emph{Stretched Far} accumulators are intended primarily for use
in understanding backgrounds.

The goal of the \emph{Stretched Window} accumulator is to include the entire
Compton pulse, while excluding low-energy background pulses and the entirety of
high-energy background bremsstrahlung pulses (which go into the \emph{Stretched
Far} accumulator).

The number of samples summed into each accumulator for each helicity window is
also read out.  This is necessary for pedestal subtraction during analysis.

\subsection{Triggered Mode} \label{sec:TriggeredMode}

\begin{figure*}[t] \begin{center} \includegraphics[bb=53 167 695
495,clip=true,width=14.0cm]{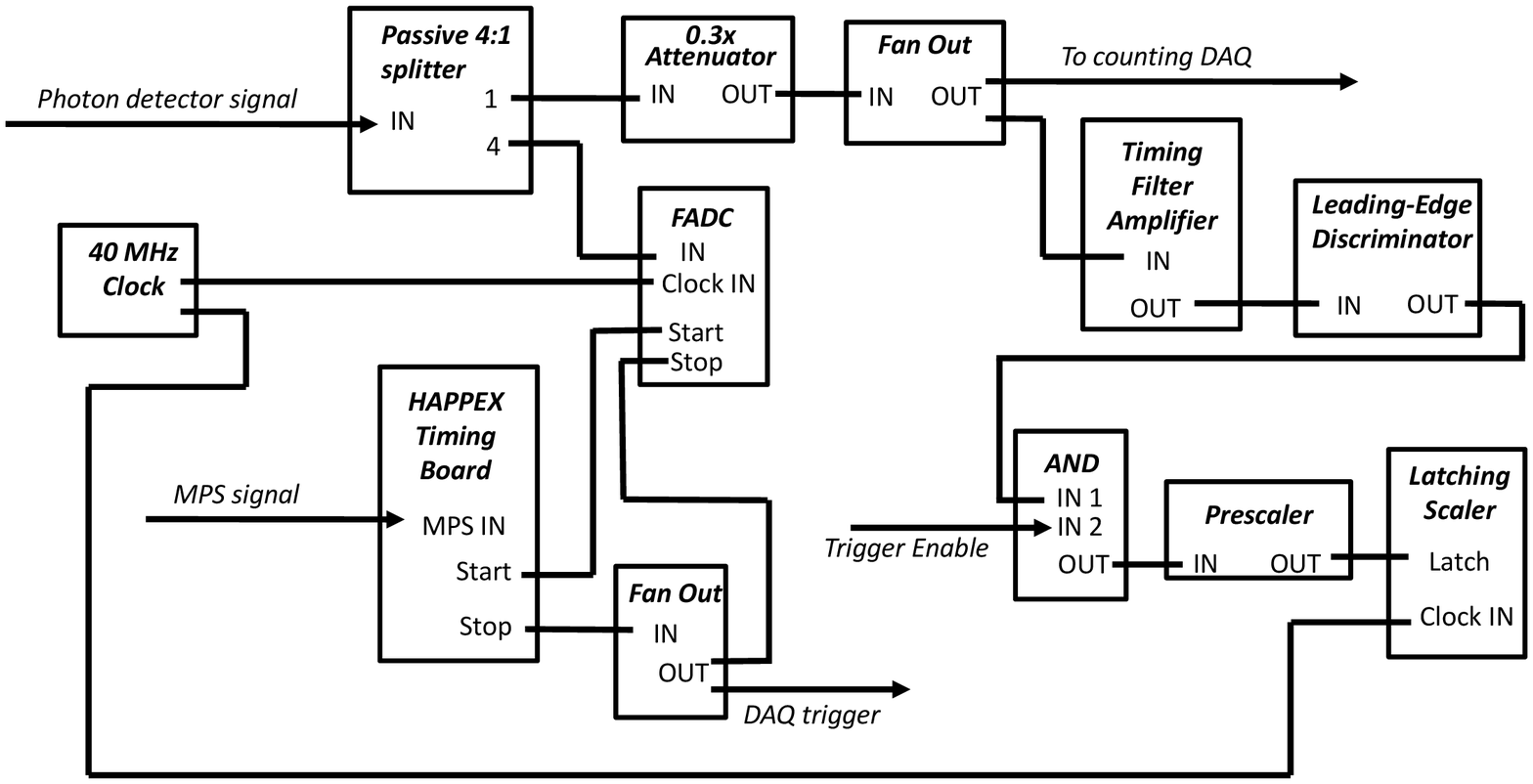} \caption{Simplified schematic
of the upgraded integrating Compton DAQ. \label{fig:daq_wiring}} \end{center}
\end{figure*}

To allow for study of individual pulse integrals and shapes, a sampled triggered
mode is implemented in parallel with the accumulator mode.   For the triggered
mode, while the detected pulse shape is continuously sampled by the FADC, the
clock times of an external trigger are recorded in a CAEN V830 latching scaler.
The latching scaler counts clock ticks and accepts an external trigger; when it
receives a trigger, it stores the current clock counter for subsequent readout.
During readout after the end of the helicity window, a programmable sampling
period, usually 500 ns, is read out from the FADC memory for each latched
trigger time, starting some programmable interval before the stored trigger
time, in order to also read out the pulse shape before the trigger.  The samples
making up the pulse can then be summed numerically (with only this sum being
written into the datastream), or all of the samples for a single trigger can be
saved to the datastream.  The readout time for either method is equivalent, but
writing out individual samples requires considerably more disk space.
Therefore, most triggered pulses are integrated, and only a few fully sampled
triggered pulses are written out for each helicity period.

There is a concern with readout time: because the data are stored in alternating
buffers, the DAQ must be finished reading out the triggered data by the end of
the subsequent helicity period.  Also, since the scaler information is not
buffered, the diagnostic scalers and all the trigger times must be read during
the short interval between \(T_{stop}\) and \(T_{start}\).  The number of
trigger times stored and the number of samples read out must therefore be
limited so that the DAQ is finished reading out before the data are overwritten.
Limits are therefore placed on the number of samples stored, and the GSO photon
trigger is prescaled (using a remotely controllable CAEN V1495, programmed to
work as a prescaler) before being sent to the latching scaler.  Prescaling
allows the latched triggers to be distributed across the helicity window, in
order to monitor any systematic signal variation as a function of time within
the helicity window.

The latching scaler must run on the same clock as the FADC, or the two sampling
rates may drift with respect to one another, causing the trigger times stored in
the latching scaler to become incorrect.  Although the CAEN V830 latching scaler
is specified to accept clocks up to 250 MHz, the NIM output signal of the
internal FADC clock was found to be unable to properly trigger the scaler at
such high rates.  The latching scaler clock input is therefore the same external
40 MHz clock used by the FADC.  There is a drawback to this method: the coarse
clock on the latching scaler causes some jitter in the trigger times.

A set of NIM logic gates, controlled via programmable output bits from the
Trigger Interface Register (TIR) of the DAQ, allows the trigger input to the
V830 to be remotely selected.  The standard trigger is, of course, the signal
from the photon detector. This is split by a 4:1 passive splitter; the majority
of the signal is sent directly to the FADC, and the rest is attenuated, sent
through a Timing Filter Amplifier (TFA) for shaping, and then to a discriminator
with a very low threshold.  A prescaled sample of the pulses which fire the
discriminator are sent to the latching scaler.  Use of a passive splitter for
the photon signal going into the FADC is important, to avoid introducing a
rate-dependent gain shift.  For example, a significant (2.5~mV) gain shift
effect was seen in the TFA when going between trigger rates with the cavity
locked compared to with the cavity unlocked with a 100~\(\mu\)A electron beam.
A schematic of the DAQ is shown in Fig.\ \ref{fig:daq_wiring}.  

A remotely programmable (nominally 1 kHz) square pulse, either in coincidence
with an LED pulser, or alone for looking at samples uncorrelated to pulses, can
also be used as a trigger.  The photon DAQ can also be triggered on the Compton
electron detector signal, allowing analysis of electron/photon coincidences. 

The standard triggered-running mode reads Compton photon detector triggers for
three helicity windows, and then reads out random samples every fourth helicity
window.  The random samples are chosen in the readout code by stepping through
the length of the helicity window.  These random samples are used for background
and pileup analysis.

\begin{figure}[] \begin{center} \includegraphics[width=8.0cm]{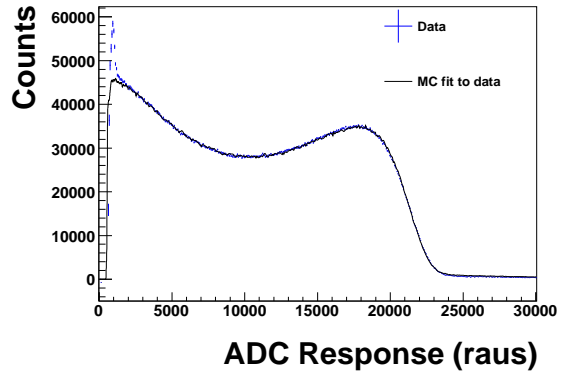}
\caption{A measured Compton photon energy spectrum. The triggered data is fit to
GEANT4 MC data with only two free parameters: a horizontal scale factor and a
vertical scale factor.  A smearing factor which accounts for PMT and DAQ
resolution, as well as GSO light collection, is also included.  The fit is good
enough that the data and MC fit are indistinguishable, except at low photon
energies, where the triggered-data background-subtraction (which is done
absolutely by taking beam current and trigger rates into account) is incorrect
due to a rate-dependent gain shift of the TFA, which causes a
trigger-discriminator threshold shift.  \label{fig:comptSpect}} \end{center}
\end{figure}

\begin{figure}[] \begin{center} \includegraphics[width=8.0cm]{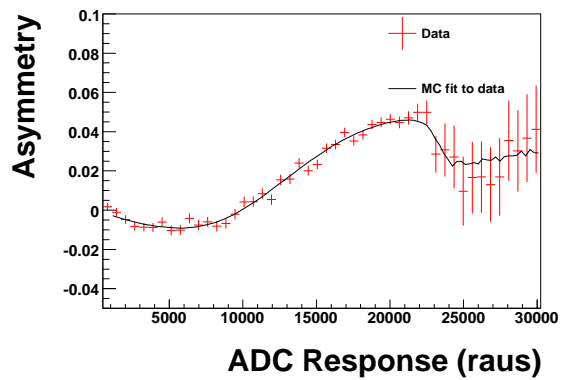}
\caption{The measured Compton asymmetry plotted against ADC response due to
energy deposited in the GSO.  The triggered data is compared to GEANT4 MC data
with no adjustable parameters.  (The horizontal scale is taken from the fit from
Fig.\ \ref{fig:comptSpect} and the vertical scale is set by the measured
\(P_{e}\) and \(P_{\gamma}\).) \label{fig:asymSpect}} \end{center} \end{figure}

A Compton photon energy spectrum, measured using the summed triggered mode of
the DAQ, is shown in Fig.\ \ref{fig:comptSpect}, where the horizontal axis is
ADC response due to energy deposited in the GSO in summed raw-ADC units (raus).
A plot of the measured Compton asymmetry as a function of ADC response is shown
in Fig.\ \ref{fig:asymSpect}.  These triggered spectra have been fit using the
shapes predicted by the GEANT4 Monte Carlo (MC) used to calculate \(A_{th}\).
This MC includes information about the electron and photon beam energies; the
detector and collimator position relative to the photon beam, as discussed in
Sec.\ \ref{sec:positioning}; the detector linearity, as discussed in Sec.\
\ref{sec:linTesting}; pileup effects; and smearing effects due to photoelectron
statistics and light collection in the detector.

\section{Integrating-Data Analysis} \label{sec:AccAnal}

Extracting an electron beam polarization from accumulator data requires making
cuts to the data based on parameters such as electron beam current and photon
cavity power.  An asymmetry can then be calculated in several different ways; an
accumulator (Sec.\ \ref{sec:threshaccs}) and method of asymmetry calculation
(Sec.\ \ref{sec:asymcalc}) must be chosen based on which measurement gives the
lowest error.  Statistical (Sec.\ \ref{sec:AsymErr}) and systematic (Sec.\
\ref{sec:SysErr}) errors on the asymmetry must then be calculated.  

Since the FADC actually stores the signal as offsets below the pedestal value,
as in Fig.\ \ref{fig:trigpulse}, the integrated signal for each window is
calculated as \begin{equation} \label{eq:accraw} S_n = N_{n}\bar{P} -
Acc_{n},\end{equation} where \(S_n\) is the physics signal extracted from the
\(n\)th FADC accumulator, \(N_{n}\) is the number of samples that have been
summed into the accumulator, \(\bar{P}\) is the best estimate of the average
pedestal value for each sample, and \(Acc_{n}\) is the integrated ADC value for
the helicity window.

The accumulator values are used to calculate the asymmetry \(A_{exp}\) from Eq.\
\ref{eq:AsymScatt}: for each period of right- or left-circular laser
polarization, separate sums of accumulator values for all positive- and
negative-helicity windows are made.  A sum is also made of accumulator values
for the adjacent cavity-unlocked periods, to determine background, \(B\), for
the cavity-locked period.  The measured asymmetry needs to take into account the
background, such that Eq.  \ref{eq:AsymScatt} becomes
\begin{equation}\label{eq:AsymWithBkg} A_{exp} = \frac{(\langle M^{+}\rangle
- \langle B\rangle) - (\langle M^{-}\rangle - \langle B\rangle)}{(\langle
  M^{+}\rangle - \langle B\rangle) + (\langle M^{-}\rangle - \langle B\rangle)},
  \end{equation} where \(\langle \rangle\) denotes the mean accumulator value
  per helicity window over each cavity (-locked or \mbox{-unlocked}) period.
  Here, 
  \(M^{+(-)}\) is the measured integrated signal plus background for positive
  (negative) helicity electrons (where \(S = M - B\)).  \(A_{exp}\) is
  calculated separately for each laser polarization.  It is assumed in this
  calculation that \(B^+ = B^-\), which is true as long as the electron beam
  parameters (such as beam position and charge) are carefully kept
  helicity-independent. This helicity independence is also necessary for making
  a Compton polarization measurement with better than 1\% systematic error, as
  well as for performing parity-violation measurements
  \cite{Paschke:ParityBeam07}.  The background then cancels in the numerator of
  Eq.\ \ref{eq:AsymWithBkg}.

Since \(N_{n}\) is always the same for every helicity window in the \emph{All}
accumulator (independent of the state of the helicity or laser cavity),
\(N_n\bar{P}\), from Eq.\ \ref{eq:accraw}, therefore cancels in both the
numerator and denominator of Eq.\ \ref{eq:AsymWithBkg}, and an \emph{All}
accumulator measurement is insensitive to the choice of pedestal value.  The
same is not true of accumulators with thresholds.

\subsection{Analysis with Threshold Accumulators} \label{sec:threshaccs}

Using the threshold accumulators (\emph{Window} or \emph{Stretched Window}
accumulators) for Compton data analysis increases signal-to-noise but comes with
an inherent additional systematic error, and therefore must be done with care:
one main advantage of making an integrating measurement is the elimination of
thresholds; when thresholds are reintroduced, these systematics return. 

Since \(V_{near}\) may be placed very close to the pedestal and the measurement
is energy-weighted, introducing a threshold does not have a large effect on
\(A_{exp}\) to first order.  However, complicated pileup effects can distort the
measured asymmetry, since small background pulses that do not cross \(V_{near}\)
when the cavity is in the unlocked state, may cross \(V_{near}\) when they pile
up with Compton photon pulses.

The main additional systematic effect that comes from using a threshold
accumulator, however, is a sensitivity to the accurate determination of the
pedestal value, \(\bar{P}\).  Since the raw accumulator data must be
pedestal-subtracted, as in Eq.  \ref{eq:accraw}, and background-subtracted, as
in Eq. \ref{eq:AsymWithBkg}, and there are a different number of samples in each
helicity window (specifically when the cavity is locked vs.\ unlocked), the
result is sensitive to the value of \(\bar{P}\).  Since the FADC pedestal is not
stable (slow drifts of the pedestal on the order of \({\sim}\)0.1 channels in
hours or \({\sim}\)0.4 channels in weeks have been observed), it is very
difficult to subtract the correct pedestal value.  Systematic errors introduced
due to a 0.4 channel pedestal uncertainty are around 0.5-1\%, depending on the
relative signal-to-background rates.  Reduction of this systematic error could
be achieved by shutting off the electron beam or detector high voltage every few
hours during data-taking, in order to monitor the pedestal, but this imposes
significant overhead, and still does not solve the problem for shorter-timescale
pedestal drift.  

Use of a threshold also introduces a second-order distortion in the measured
energy-weighted asymmetry, since \(V_{near}\) discards a larger fraction of each
lower-energy photon pulse which crosses it compared to higher-energy pulses in
the \emph{Window} accumulator.  The \emph{Stretched Window} accumulator is more
complicated, since it opens a window which integrates \(N_4^{\mathit{after}}\)
samples after the signal re-crosses the threshold, and the timing of the
threshold-crossing walks depending on the photon energy (again causing a
distortion in the energy-weighted asymmetry).  Because of the fast rise-time of
each photon pulse, this is not a problem for the initial threshold crossing in
the \emph{Stretched Window} accumulator.  To improve understanding of the
\emph{Stretched Window} accumulator data and facilitate better extraction of a
polarization from this data, a future version of the SIS3320 firmware would stop
counting \(N_4^{\mathit{after}}\) after \(V_{near}\) was crossed the first time,
instead of the second time, thereby integrating the same number of samples for
each pulse, independent of the pulse height.  The same would be done for the
\emph{Stretched Far} accumulator and \(N_5^{\mathit{after}}\).

These effects cause a non-negligible systematic difference in the measured
asymmetry for each accumulator (e.g.\ the measured asymmetry for HAPPEX-III is
systematically 0.3\% higher in the \emph{Stretched Window} accumulator relative
to the \emph{All} accumulator and is 1.1\% higher in the \emph{Window}
accumulator relative to the \emph{All} accumulator), and must therefore be taken
into account when calculating \(A_{th}\).  

The use of the \emph{All} accumulator was therefore found to produce results
with smaller overall systematic errors, since it is better understood.

\subsection{Asymmetry Calculation} \label{sec:asymcalc}

\begin{figure*}[tH!] \begin{center}
\subfloat[]{\label{fig:slugright}\includegraphics[width=8cm]{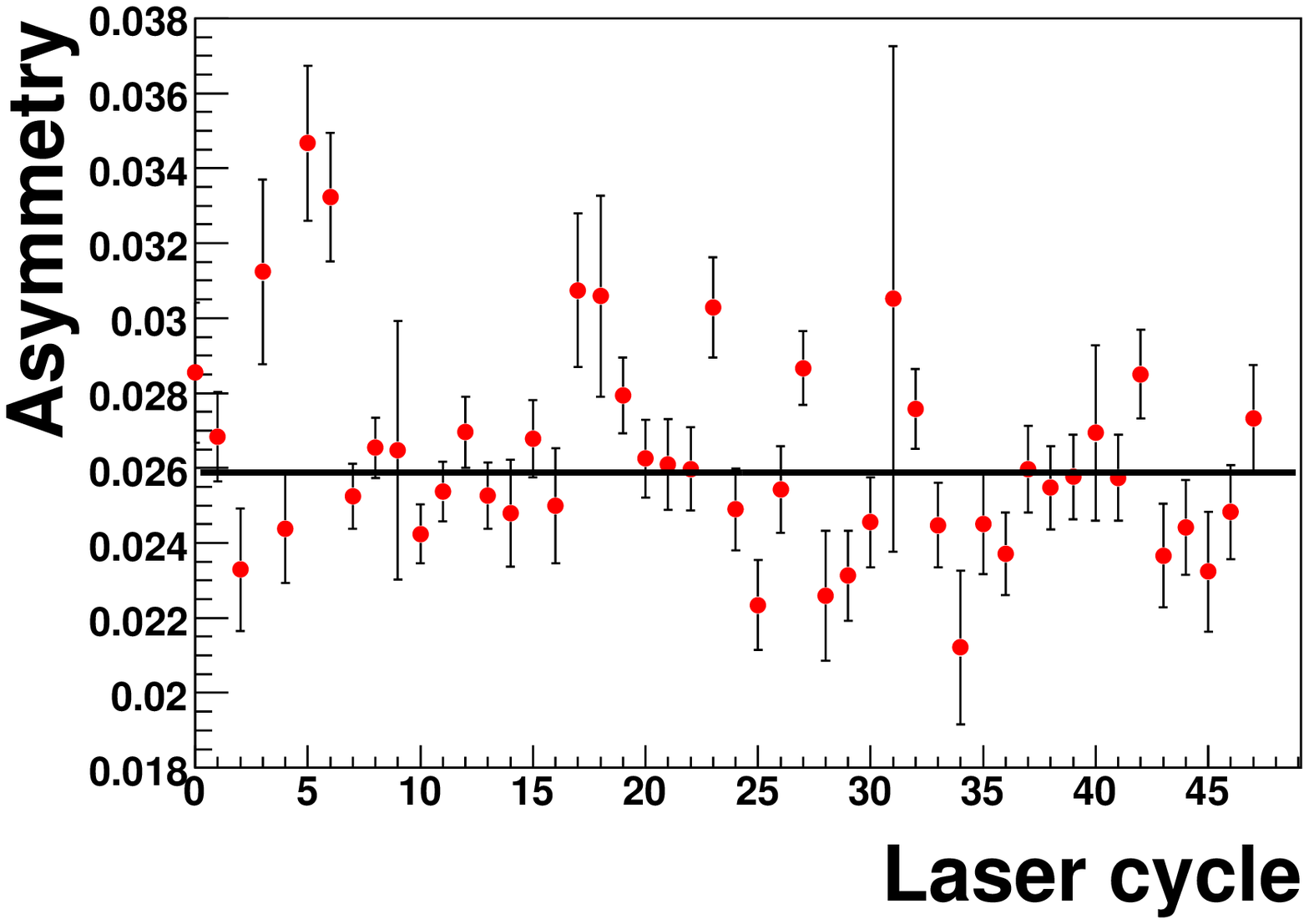}}
\subfloat[]{\label{fig:slugleft}\includegraphics[width=8cm]{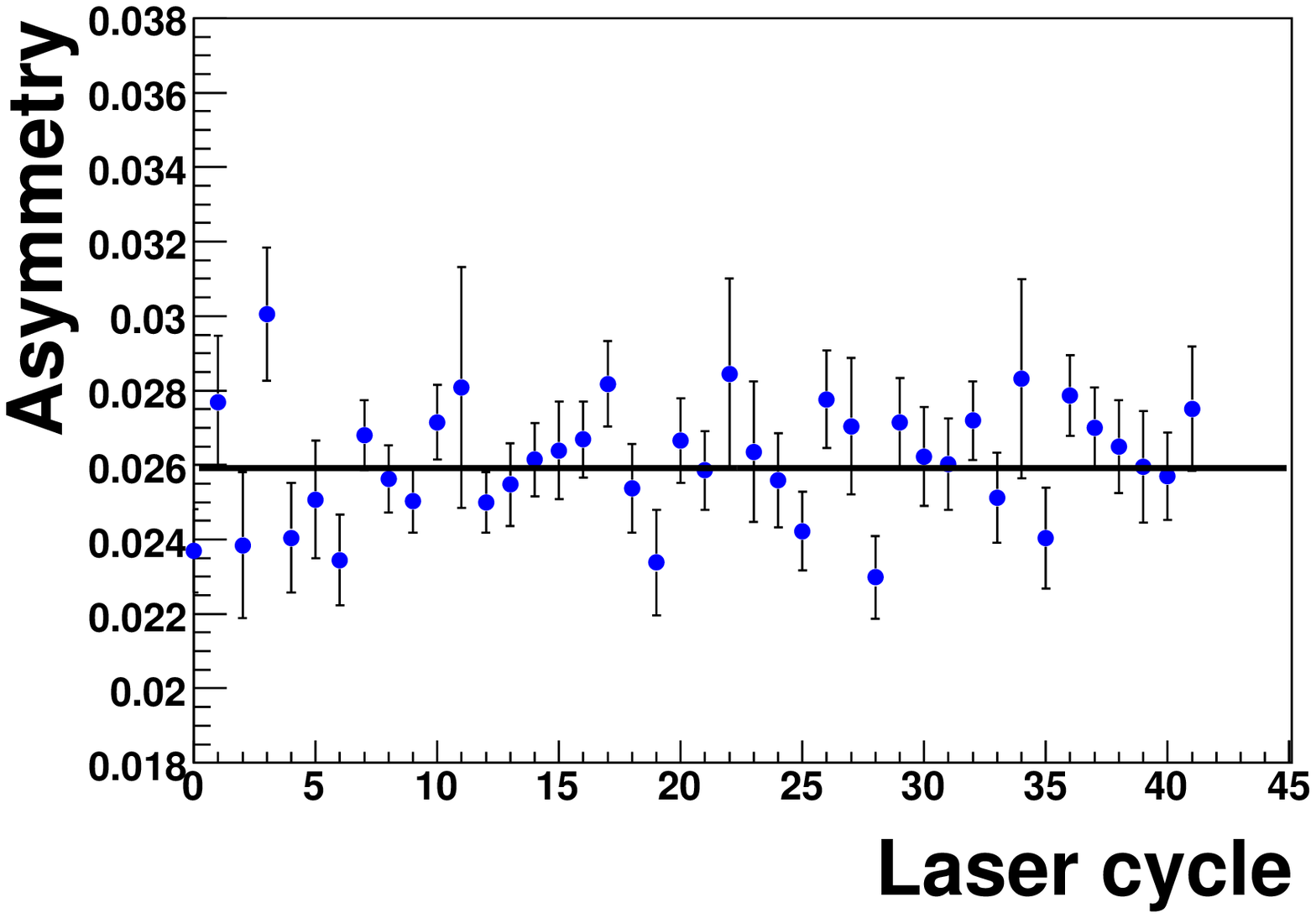}}
\caption{A typical Compton slug for (a) laser-right and (b) laser-left, where
the absolute value of the asymmetry has been taken.  Each data point is a
separate laser-cycle including local background subtraction.  Error bars are
statistical as defined in Sec.\ \ref{sec:AsymErr}.  The solid line is a constant
fit to the data.  \label{fig:slug60}} \end{center} \end{figure*}

There are several options for extracting an asymmetry (calculated as in Eq.
\ref{eq:AsymWithBkg}) from the Compton accumulator (\emph{All}, \emph{Window},
or \emph{Stretched Window}) data.  Three options discussed here are called
laser-wise, for which an asymmetry is calculated for each laser cycle; run-wise,
for which an asymmetry is calculated for each one- or two-hour long run; and
pair-wise, for which an asymmetry is calculated for each helicity pair.

The laser-wise method of extracting an asymmetry involves calculating a separate
mean of the accumulated value (averaged over the number of helicity windows) of
all helicity-plus and -minus windows for each cavity-locked period, as described
in Sec.\ \ref{sec:AccAnal}.  A mean local background from the two
cavity-unlocked periods adjacent to the cavity-locked period is also calculated.
Since the background contributions to signal fluctuate quickly and tend to drift
on the timescale of minutes, local background determination is advantageous.  A
separate statistical error bar is then assigned for each laser-cycle point (as
described in Sec.\ \ref{sec:AsymErr}), and the points are collected into
\({\sim}\)50 laser-cycle-long ``slugs'' of data (broken up so that no run is
divided between multiple slugs and each slug contains two long runs or several
short runs with no configuration changes).  Measured asymmetries from a typical
slug are shown in Fig.\ \ref{fig:slug60}.  The mean for each slug is taken as a
separate data point (as plotted in Sec.\ \ref{sec:results}).  This method
appears to have the best balance of consistency checks and resistance to
excessive noise, and was used to determine the electron beam polarization for
the HAPPEX-III measurement.

\begin{figure*}[t] \begin{center}
\subfloat[]{\label{fig:diff}\includegraphics[width=8cm]{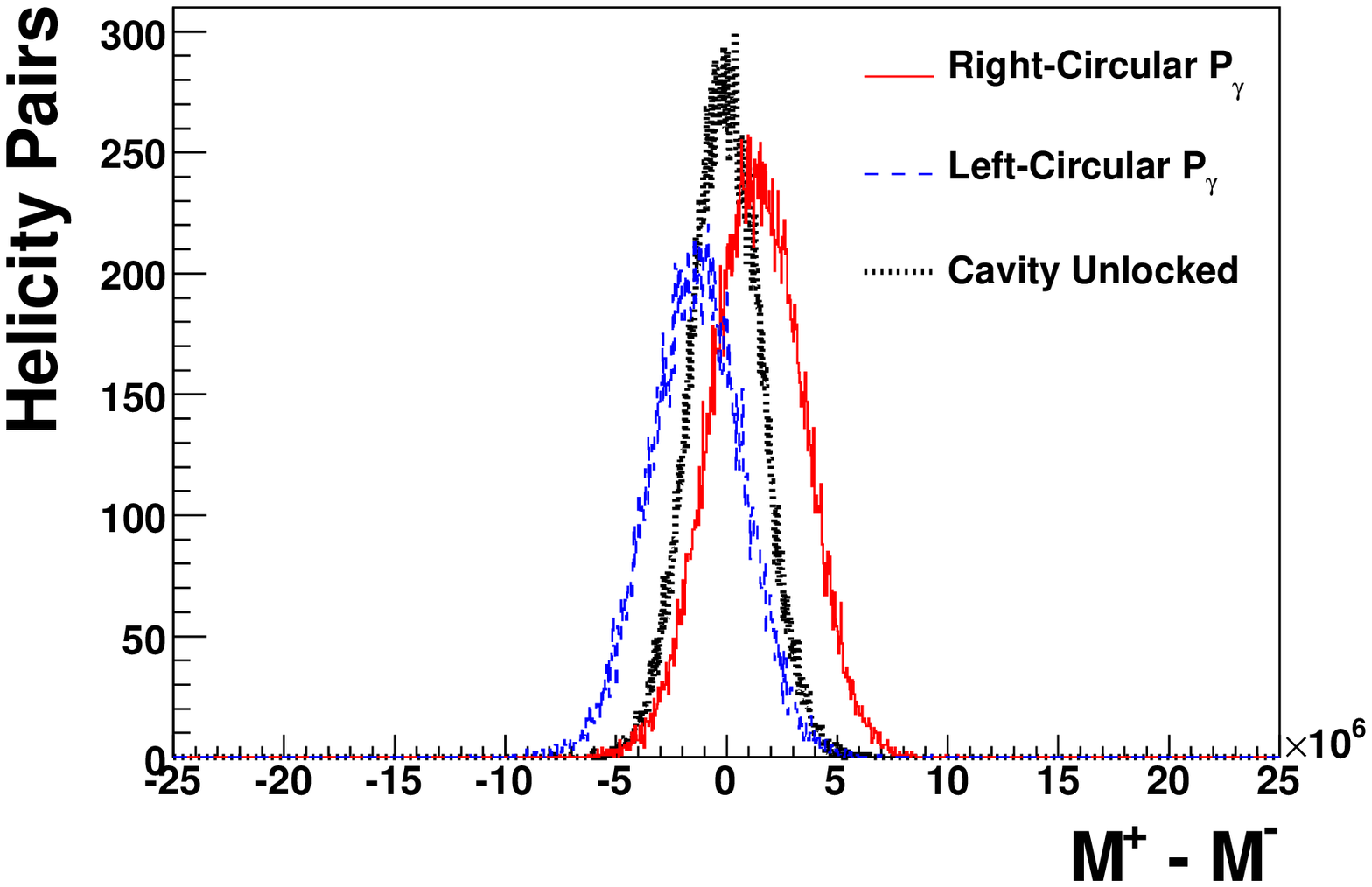}}
\subfloat[]{\label{fig:sum}\includegraphics[width=8cm]{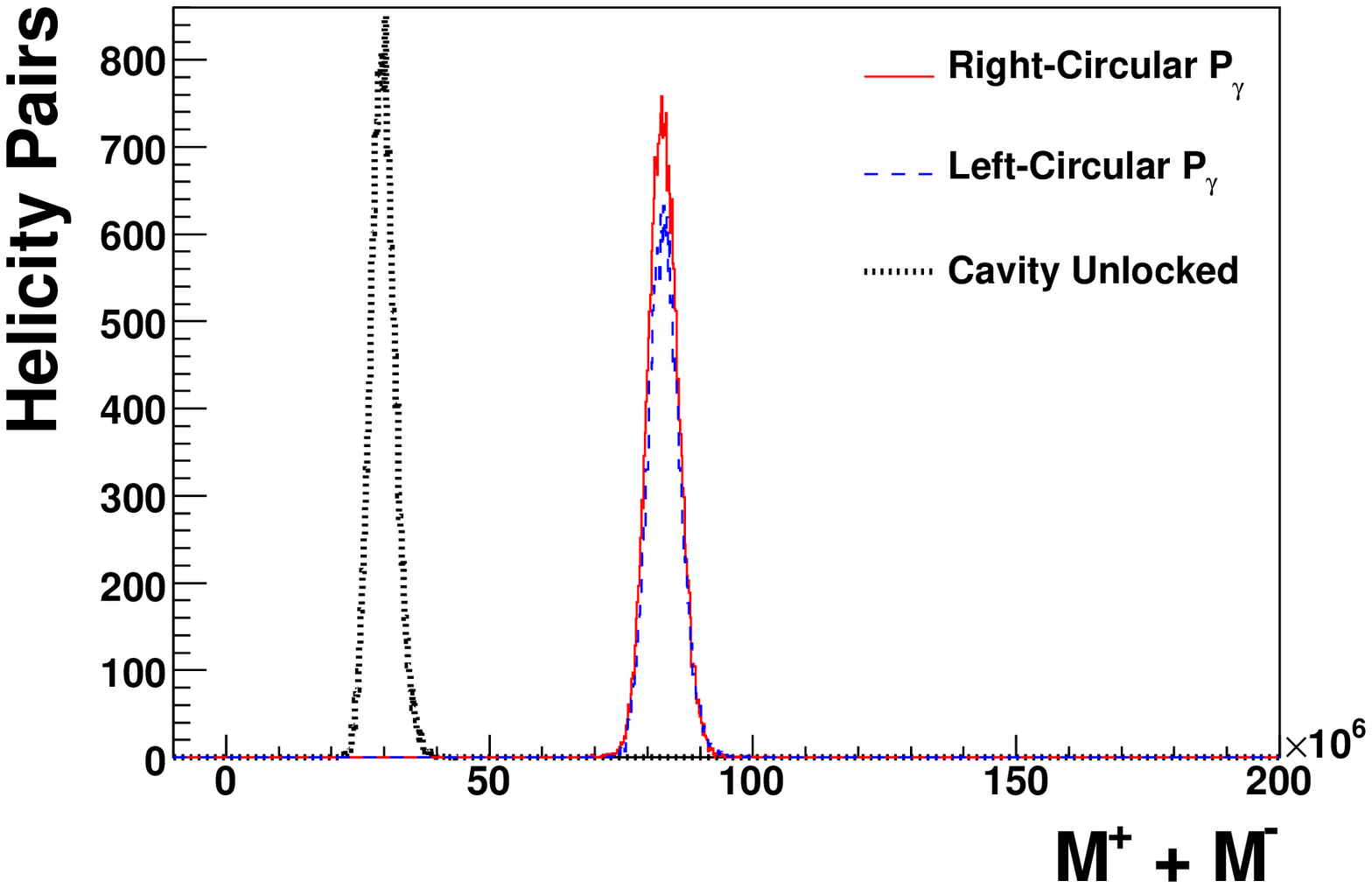}}
\caption{Histograms of the (non-background-subtracted) (a) numerator and (b)
denominator of the Compton asymmetry for an entire two hour long run.
\label{fig:sumanddiff}} \end{center} \end{figure*}

The data can also easily be broken up into runs: a sum and a difference for each
(\({+}{-}\)) helicity pair in the run is calculated (separately for each laser
polarization, of course), and the mean of these values is taken over the entire
run.  A mean background value for the entire run is also calculated.  These
numbers are then used as the numerator and denominator in Eq.\
\ref{eq:AsymWithBkg}.  Histograms of the sum, difference, and background for a
typical run are shown in Fig.\ \ref{fig:sumanddiff}.  This run-wise method of
calculating the asymmetry is particularly useful for running at lower rates:
since each (\({\sim}\)90 s) laser-cycle has low photon statistics at low rates,
doing a laser-wise analysis is impossible.  This method has the disadvantage
that the background level is averaged over the entire run, which significantly
increases the error due to background subtraction when backgrounds are unstable.

\begin{figure}[] \begin{center} \includegraphics[width=8.0cm]{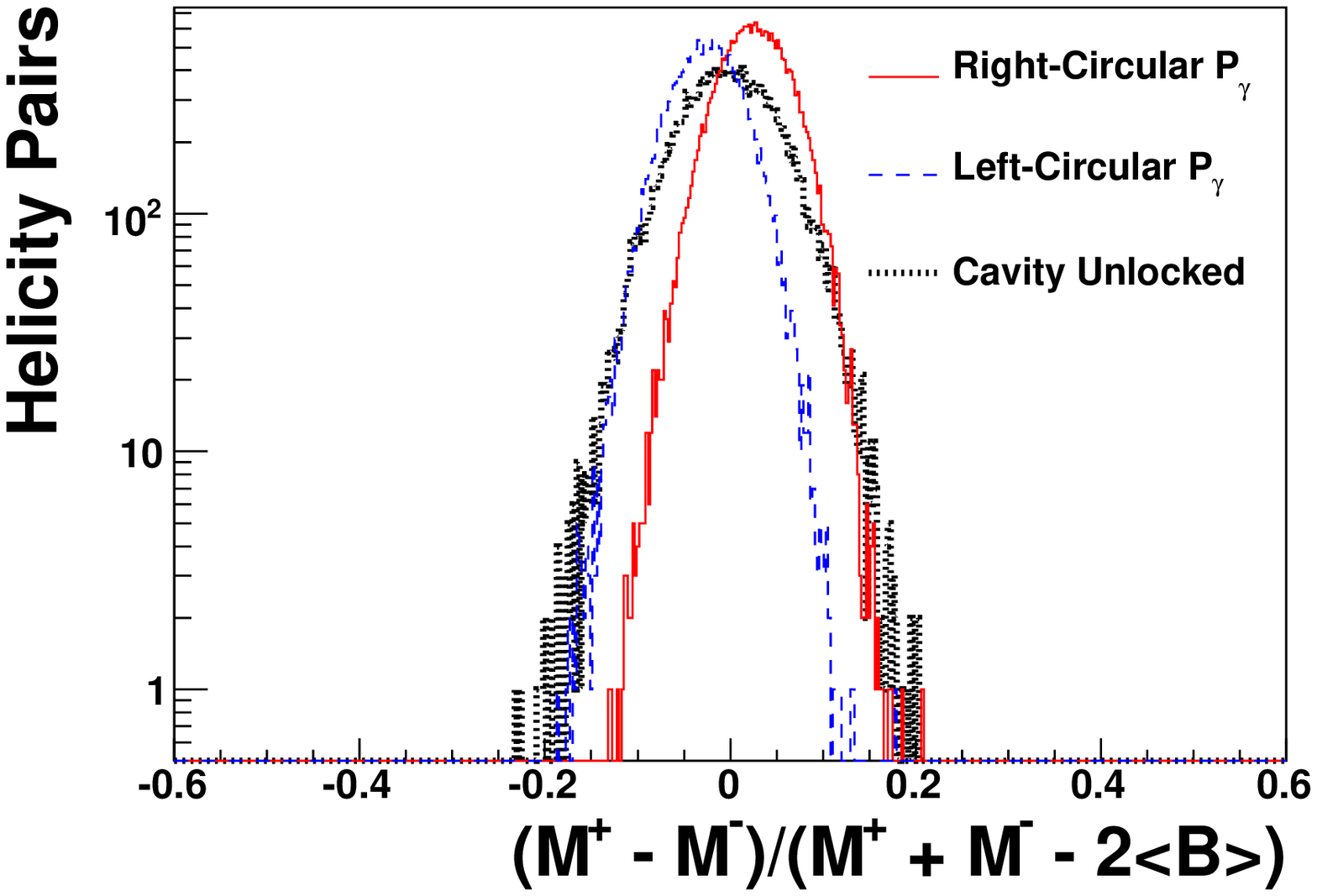}
\caption{Histogram of a background-subtracted Compton asymmetry taken for every
pair in a single two hour long run. \label{fig:pairwiseasym}} \end{center}
\end{figure}

The pair-wise method involves calculating a separate (background-subtracted)
asymmetry, as shown in Fig.\ \ref{fig:pairwiseasym},  for each helicity pair.
Preliminary results using this method are described by Parno et al.\
\cite{Parno:INPC10}.
The simplest implementation of this method also uses a background calculated for
the entire run, and so a run-averaged background value is subtracted rather than
a local one.  Calculating an error bar for this method requires use of not only
the statistical error on the distribution from Fig.\ \ref{fig:pairwiseasym}, but
also the statistical error on the background spectrum.  Unfortunately, when the
background is high and unstable (with an integrated background to signal ratio
of more than \({\sim}\)1 or a background distribution RMS width of more than
\({\sim}10\%\) of the mean), the pair-wise distribution becomes non-Gaussian,
and therefore this method does not give correct statistical errors nor mean
asymmetries.

\subsubsection{Calculation of Statistical Errors} \label{sec:AsymErr}

To assign statistical error bars when calculating a run-wise or laser-wise
asymmetry, the RMS width of each sum, difference, and background distribution
(as shown in Fig.\ \ref{fig:sumanddiff} for a whole run, but done separately for
each laser-cycle for a laser-wise analysis) is divided by the square root of the
number of data points.  This quantity for the sum, difference, and twice the
background is labeled \(\sigma_S\), \(\sigma_D\), and \(\sigma_B\) respectively.
These values are then used to calculate a statistical error on each point,
treating \(\langle M^+-M^-\rangle \), \(\langle M^++M^-\rangle \), and \(\langle
2B\rangle \) as independent variables in Eq.\ \ref{eq:AsymWithBkg}: 
\begin{align}\sigma^2 = &\frac{\sigma_D^2}{(\langle M^+\rangle +\langle
M^-\rangle -2\langle B\rangle )^2} + \nonumber \\
&\frac{(\sigma_S^2+\sigma_B^2)^2(\langle M^+\rangle -\langle
M^-\rangle)^2}{(\langle M^+\rangle +\langle M^-\rangle -2\langle B\rangle )^4}.
\end{align}
This method of error calculation allows the width of the background distribution
to be properly taken into account when assigning errors. 

A table of the average \emph{All} accumulator sum and difference values and the
RMS widths of the distributions measured during the HAPPEX-III experiment is
given in Sec.\ \ref{sec:results}.

\subsection{Calculation of Systematic Errors on the \emph{All} Accumulator}
\label{sec:SysErr}

As long as the electron beam parameters are kept minimally helicity dependent,
the main systematic error on an \emph{All} accumulator integrating Compton
asymmetry is due to the observed PMT gain shift between cavity-locked and
-unlocked states.  The systematic error due to the gain shift results from
uncertainty in the size of the gain shift itself (the signal-to-background ratio
during a three-month-long experiment does not stay constant, and the gain shift
changes depending on the relative rates).  There is also an error on the gain
shift due to pedestal uncertainty, since a gain shift correction is implemented
(by scaling \(\langle B\rangle \) in Eq.\ \ref{eq:AsymWithBkg}) after pedestal
subtraction, and is therefore sensitive to the correct assignment of the
pedestal value, which no longer cancels exactly, even for the \emph{All}
accumulator.  

A pedestal shift between cavity-locked and -unlocked states would also be a
source of systematic error, but there is no such observed pedestal shift in this
setup.  

Systematic errors in the analyzing power must also be calculated, and these are
estimated by changing the beamline and electron beam parameters input into the
GEANT4 MC, e.g.\ the photon beam position on the collimator or the electron beam
energy, over the experimentally possible range of values, and measuring the
fractional change in \(A_{th}\).  There is also a systematic error on \(A_{th}\)
due to detector non-linearity: the detector linearity is measured as described
in Sec.\ \ref{sec:linTesting}, and this measured value is input into the MC.
The systematic error on \(A_{th}\) due to non-linearities is then estimated by
tweaking the non-linearity input into the MC and monitoring the effect of these
changes on the fit of the MC data to the Compton spectrum, as in Fig.\
\ref{fig:comptSpect}, as well as the effect on \(A_{th}\).  

A table of systematic errors on the Compton integrating measurement for the
HAPPEX-III experiment is given in Sec.\ \ref{sec:results}.

\section{Measurement Results}\label{sec:results}

The upgraded photon arm of the Compton polarimeter was used to continuously
measure the CEBAF electron beam polarization during several experiments,
including HAPPEX-III, which ran in 2009.  This particular measurement achieved a
0.94\% relative (0.84\% absolute) systematic error, which mostly came from an
uncertainty in laser photon polarization in the cavity.  As discussed in Sec.\
\ref{sec:Apparatus}, the photon polarization is determined by combining an
on-line measurement of the polarization at the cavity exit with a measurement of
the cavity transfer function; there is a 0.8\% error on the cavity transfer
function measurement.  The systematic errors on the asymmetry measurement are
detailed in Sec.\ \ref{sec:SysErr}.  The breakdown of Compton systematic errors
is shown in Table \ref{tab:syserrs}.

\begin{table} \begin{center} \begin{tabular}{  l  l  } \hline
\multicolumn{2}{c}{\textbf{Systematic Errors}} \\ \hline \textbf{Laser
Polarization} & 0.80\% \\ \hline \textbf{Analyzing Power:} & \\ Non-linearity &
0.3\% \\ Electron Energy Uncertainty & 0.1\% \\ Collimator Position & 0.05\% \\
MC Statistics & 0.07\% \\ \hline \textbf{Total on Analyzing Power} & 0.33\% \\
\hline \textbf{Gain Shift:} & \\ Background Uncertainty & 0.31\% \\ Pedestal
Uncertainty & 0.20\% \\ \hline \textbf{Total on Gain Shift} & 0.37\% \\ \hline
\textbf{Total} & \textbf{0.94}\% \\ \hline \end{tabular} \end{center}
\caption{Breakdown of Compton systematic errors using the \emph{All} accumulator
during HAPPEX-III.  \label{tab:syserrs}} \end{table}

The statistical error after three months of HAPPEX-III running with the
integrating Compton DAQ was 0.06\%.  Statistical error calculation is discussed
in Sec.\ \ref{sec:AsymErr}, and a table of the measured means and RMS widths of
the (non-background-subtracted) sum and difference distributions from the
\emph{All} accumulator for each run, averaged over all of the HAPPEX-III runs,
is given in Table \ref{tab:RMSwidths}.  The means and RMS widths of the
background and pedestal are also given.

\begin{table} \begin{center} \begin{tabular}{ p{2.3cm}  p{1.8cm}  p{2.2cm} }
\hline \textbf{Measurement} & \textbf{Mean (raus)} & \textbf{RMS Width (raus)}
\\ \hline \textbf{Sum} & \(124\times10^6\) & \(13\times10^6\) \\ \hline
\textbf{Difference} & \(1.8\times10^6\) & \(5.5\times10^6\)\\ \hline
\textbf{Background} & \(54\times10^6\)& \(16\times10^6\) \\ \hline
\textbf{Pedestal} & \(24000\times10^6\) & \(0.71\times10^6\) \\ \hline
\end{tabular} \end{center} \caption{The measured means and RMS widths of the
non-background-subtracted sum and difference distributions from the \emph{All}
accumulator for each run, as in Fig.\ \ref{fig:sumanddiff}, averaged over all of
the HAPPEX-III runs (with \({\sim}90\times 10^3\) MPSs in each run).   The means
and RMS widths of the background (summed over pairs of windows) and pedestal
(for single 33~ms helicity windows) are also given.  \label{tab:RMSwidths}}
\end{table}

Because there were gaps in the run period during which the polarization was not
monitored by the Compton polarimeter (due in part to electron beam instability
which made Compton measurements impossible and in part to a period of time for
which the laser polarization became unknown due to an equipment failure), an
additional error of 0.2\% was included.  

\begin{figure*}[tH!] \begin{center} \includegraphics[bb=70 0 515 165,
width=16.0cm]{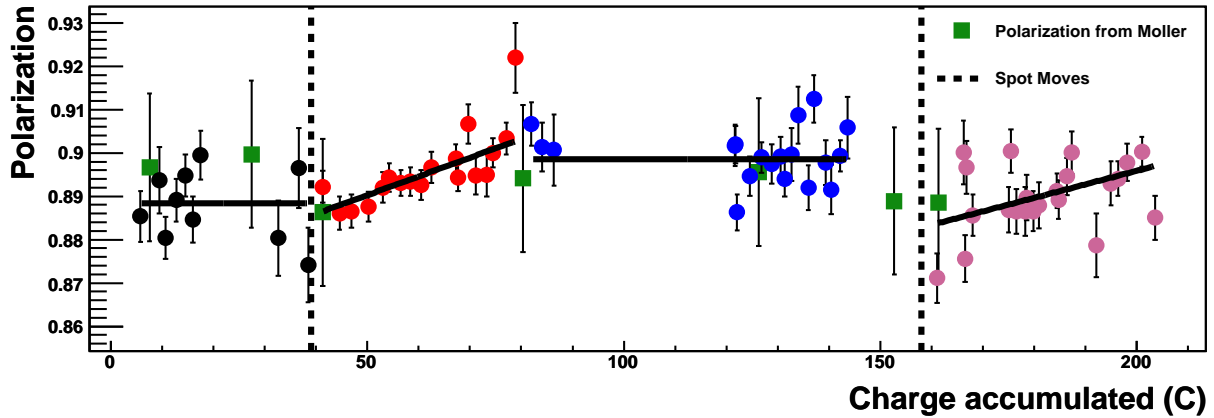} \caption{A plot of measured polarization
vs.\ HAPPEX-III charge accumulated.  The round points are measured using the
Compton polarimeter integrating DAQ, and error bars are statistical only.  The
vertical dashed lines mark when the laser spot was moved at the accelerator
source, and the solid lines are linear fits to the Compton data, where the
HAPPEX-III run period is broken into four distinct polarization periods and fit
accordingly.  The square points are from the Hall A M{\o}ller polarimeter
\cite{Dale:Moller97}, and the error bars on these points include a 1.7\%
systematic error.  \label{fig:PolvsCh}} \end{center} \end{figure*}

A plot of the electron polarization as a function of HAPPEX-III incident beam
charge accumulated is shown in Fig.\ \ref{fig:PolvsCh}.  In this plot, the
vertical dashed lines mark when the laser spot was moved at the accelerator
source; there is a distinct measured change in polarization behavior following
spot moves \cite{Sinclair:PolarizedSource07}.  The observed gradual increase in
polarization following spot moves is consistent with a gradual degradation of
the photocathode surface, which has been observed to increase electron
polarization while the quantum efficiency drops
\cite{Tang:SLACSource95,Saez:Source97}.  The photon arm of the Compton
polarimeter has measured the average beam polarization over the
HAPPEX-III run to be \([89.41 \pm 0.05(\text{stat}) \pm 0.84(\text{sys}) \pm
0.18(\text{gaps})]\%\).

\section{Conclusion} 

The photon arm of the Jefferson Lab Hall A Compton polarimeter has been upgraded
with a new integrating DAQ and photon detector.  These have been used to
determine the beam polarization during HAPPEX-III with better than 1\% total
error at 100~\(\mu\)A electron beam current and 3.4~GeV electron beam energy.
These results agree with concurrently running M{\o}ller polarimeter
measurements, but have higher precision.  Polarization measurements made using
the upgraded Compton DAQ and photon detector also show a marked improvement
compared to those made with the original apparatus, with an overall error
improved by about a factor of three.

The upgraded Compton photon arm has been used in several experiments which have
run in Hall A since 2009 in addition to HAPPEX-III: d\(_2^n\)
\cite{d2nProposal}, PVDIS \cite{PVDISProposal}, PREX \cite{PREXProposal}, and
DVCS \cite{DVCSProposal1}.  Future Hall A parity experiments, such as the MOLLER
\cite{MollerProposal} and SoLID \cite{SOLIDProposal} experiments, also require
electron beam polarimetry with sub-1\% precision and will use an integrating
Compton photon DAQ.  A similar integrating DAQ was also installed in the new
Hall C Compton polarimeter at Jefferson Lab in 2010 for use in the \(Q_{weak}\)
experiment \cite{QWEAKProposal}.

Results obtained using the upgraded photon arm of the Hall A Compton polarimeter
can be compared to results obtained with Compton polarimeters used at lower
electron beam energies at NIKHEF \cite{Passchier:NIKHEFCompton98} and MAMI
\cite{Diefenbach:MAMICompton07} (which quote 2.7\% and 2\% absolute systematic
errors respectively), and at higher electron beam energies at HERA
\cite{Beckmann:HERACompton02} and SLAC \cite{Woods:SLACPol96} (which quote 1.6\%
relative and 0.5\% absolute systematic errors respectively). 

\section*{Acknowledgments} 

The authors would like to thank the Hall A technical staff for their hard work
and help in installing the upgraded beamline and calorimeter, as well as the
Jefferson Lab accelerator experts and operators for their assistance in
achieving and maintaining the high electron beam quality necessary for running
the Compton polarimeter.  

This work was supported by DOE grants DE-AC05-06OR23177, under which Jefferson
Science Associates, LLC, operate Jefferson Lab, and DE-FG02-87ER40315.

\bibliographystyle{h-elsevier} 
\bibliography{bibliography}

\end{document}